\documentclass[aps,prd,eqsecnum,onecolumn,showpacs,nofootinbib,preprintnumbers]{revtex4}
\usepackage{amsmath}
\usepackage{amssymb}
\usepackage{amsfonts}
\usepackage{latexsym}
\usepackage{graphicx}
\usepackage{dcolumn}
\usepackage{bm}
\usepackage{empheq}
\usepackage{color}
\usepackage{times}  
\def\be{\begin{equation}}
\def\ee{\end{equation}}
\def\bea{\begin{eqnarray}}
\def\eea{\end{eqnarray}}

\newcommand{\eq}[1]{\begin{equation}#1\end{equation}}

\newcommand{\ea}[1]{\begin{equation}\begin{aligned}#1\end{aligned}\end{equation}}

\newcommand{\od}[2]{\frac{\textrm{d} #1}{\textrm{d} #2}}  
\newcommand{\lrp}[1]{\left( #1 \right)}  
\newcommand{\lrsb}[1]{\left[ #1 \right]}  
\newcommand{\lrcb}[1]{\left\{ #1 \right\}}  
\newcommand{\lrmb}[1]{\left| #1 \right| } 

\def\rd{\partial}
\def\vx{\bm{x}}

\definecolor{gxhighlight}{rgb}{1,1,0.4}

\def\dep{\delta\varphi}
\def\dop{\dot{\varphi}}

\def\doq{\dot{Q}}

\begin{document}

\title{Cosmological Perturbations in Ho\v{r}ava-Lifshitz Gravity}

\author{Xian Gao$^{1)}$\footnote{gaoxian@itp.ac.cn},
 Yi Wang$^{1)}$\footnote{wangyi@itp.ac.cn},
  R. Brandenberger$^{2,3)}$\footnote{rhb@mx0.hep.physics.mcgill.ca} and A. Riotto$^{3,4)}$\footnote{Antonio.Riotto@cern.ch}}

\affiliation{1) Key Laboratory of Frontiers in Theoretical Physics,\\
     Kavli Institute for Theoretical Physics China, Chinese Academy of
    Sciences\\
    No.55, Zhong-Guan-Cun East Road, Hai-Dian District, Beijing 100190,
    P.R.China}

\affiliation{2) Department of Physics, McGill University,
Montr\'eal, QC, H3A 2T8, Canada}

\affiliation{3) Theory Division, CERN, CH-1211 Gen\`eve 23, Switzerland}

\affiliation{4) INFN, Sezione di Padova, Via Marzolo 8, I-35131 Padua, Italy}



\keywords{Cosmological perturbation theory, Inflation, Cosmology of
theories beyond the SM, Physics of the early universe}

\begin{abstract}
We study cosmological perturbations in Ho\v{r}ava-Lifshitz Gravity,
a recently proposed potentially ultraviolet-complete quantum theory
of gravity. We consider scalar metric fluctuations about a
homogeneous and isotropic space-time. Starting from the most general
metric, we work out the complete second order action for the
perturbations. We then make use of the residual gauge invariance and
of the constraint equations to reduce the number of dynamical
degrees of freedom. At first glance, it appears that there is an
extra scalar metric degree of freedom. However, introducing the
Sasaki-Mukhanov variable, the combination of spatial metric
fluctuation and matter inhomogeneity for which the action
in General Relativity has canonical form, we find that this variable has the
standard time derivative term in the second order action, and that the extra degree of
freedom is non-dynamical. The limit $\lambda \rightarrow 1$ is
well-behaved, unlike what is obtained when performing incomplete
analyses of cosmological fluctuations. Thus, there is no strong
coupling problem for Ho\v{r}ava-Lifshitz gravity when considering
cosmological solutions. We also compute the spectrum of cosmological
perturbations. If the potential in the action is taken to be of
``detailed balance" form, we find a cancelation of the highest
derivative terms in the action for the curvature fluctuations. As a
consequence, the initial spectrum of perturbations will not be
scale-invariant in a general spacetime background, in contrast to
what happens when considering Ho\v{r}ava-Lifshitz matter leaving the
gravitational sector unperturbed. However, if we break the detailed
balance condition, then the initial spectrum of curvature
fluctuations is indeed scale-invariant on ultraviolet scales. As an
application, we consider fluctuations in an inflationary background
and draw connections with the ``trans-Planckian problem" for
cosmological perturbations. In the special case in which the
potential term in the action is of detailed balance form and in
which $\lambda = 1$, the equation of motion for cosmological
perturbations in the far UV takes the same form as in GR. However,
in general the equation of motion is characterized by a modified
dispersion relation.
\end{abstract}

\maketitle \vspace{-0.5cm} \tableofcontents 

\section{Introduction}

Recently, Ho\v{r}ava proposed \cite{Horava0,Horava1}
(see also \cite{Kluson}) a model for quantum
gravity which is power-counting renormalizable and hence
potentially ultra-violet (UV) complete. The model is based on a scaling
symmetry which treats space and time differently. Hence, the model
explicitly breaks Lorentz (and hence also general coordinate)
invariance. The action chosen by Ho\v{r}ava is
power-counting renormalizable with respect to the scaling symmetry
\footnote{See also \cite{Reffert} for a recent study of the renormalizability issue.}.
For a particular value of one of the coefficients in the Lagrangian
($\lambda = 1$), the infrared (IR) limit of the action reduces to that of Einstein gravity
\footnote{See also \cite{Nastase} for a study of the IR limit.}.

Since it was proposed Ho\v{r}ava-Lifshitz (HL) gravity, as this
model is now called, has attracted a lot of attention (for a
complete list of references, the reader is referred in the recent
paper \cite{Visser2}). We will only mention some papers relevant to
our study. Initially, gravitational wave solutions in HL cosmology
were studied \cite{Soda}. The first papers on the early universe
cosmology of HL gravity are \cite{Calcagni,Kiritsis} where it was
realized that the analogs of the Friedmann equations in HL gravity
include a term which scales as dark radiation and contributes
negatively to the energy density. Thus, it is possible in principle
to obtain a nonsingular cosmological evolution with the Big Bang of
Standard and Inflationary Cosmology replaced by a bounce. In this
context, it becomes possible \cite{RHB} to provide a realization of
the ``matter bounce" alternative \cite{matterbounce} to cosmological
inflation for explaining the origin of an almost scale-invariant
spectrum of cosmological perturbations. As realized originally in
\cite{Horava1} the different ultraviolet behavior of the theory
might provide an alternative to cosmological inflation for solving
the problems of Standard Cosmology such as the horizon and flatness
problems \cite{Kiritsis}. The specific UV scaling of HL gravity
could change the usual arguments for the origin of the
scale-invariance of cosmological perturbations in inflationary
cosmology, as pointed out in \cite{Calcagni}. In fact, the dominant
UV terms in the action may lead to the possibility of obtaining a
scale-invariant spectrum of cosmological perturbations in the
expanding phase of HL cosmology without inflation \cite{Mukh} (see
also \cite{Piao}). To substantiate these conclusions, however, a
careful analysis of the theory of cosmological perturbations in HL
gravity is required, and this is the topic of the current paper.

A possibly more important reason to study cosmological perturbations
in HL gravity are basic consistency issues of HL gravity itself
\footnote{One of the authors (RB) thanks G. Dvali and E. Witten for
emphasizing this issue, and K. Zarembo for interesting discussions
on this point.}. For a particular value of one of the parameters in
the HL action, namely $\lambda = 1$, the theory has its IR fixed
point the action of GR. HL gravity has the same dynamical degrees of
freedom as General Relativity (GR), but it does not have the
complete diffeomorphism invariance of GR. Spatial diffeomorphisms
are still a symmetry, but space-dependent time reparameterizations
are no longer allowed. Thus, one loses one out of the four gauge
modes of GR, and hence one extra physical mode is expected survive.
This fact was already pointed out in \cite{Horava0} and has been
further discussed in \cite{Cai,Gao,Chen,Visser2}. This extra
physical mode, if dynamical, would lead to severe problems for HL
gravity, since no effects of extra gravitational degrees of freedom
have been observed, and since there are stringent limits on the
presence of such degrees of freedom. In \cite{Cai,Chen},
cosmological fluctuations in the absence of matter were considered.
A new physical scalar gravitational mode was found. In the limit
$\lambda = 1$, this mode was claimed to be non-dynamical \cite{Cai},
although the constraint equation in both \cite{Cai} and \cite{Chen}
showed a singularity in this limit. Perturbations in the presence of
matter were very recently considered in \cite{Saffin}, where it was
claimed that the extra gravitational degree of freedom is physical
and becomes strongly coupled in the limit $\lambda = 1$ in which GR
is to be recovered.

These differing claims form our second motivation to perform a
careful study of cosmological perturbations in the presence of
matter in HL gravity. It is crucial to perturb about a dynamical
Friedmann universe as opposed to perturbing about Minkowski
space-time \footnote{Doing the latter is inconsistent with the
background constraint equations since the presence of any matter
will lead to a non-vanishing average energy density and hence to
cosmological expansion.}. Our work builds on the paper \cite{Gao} in
which the groundwork for the present study was provided. We start
with the general metric including scalar metric fluctuations. We use
the spatial diffeomorphism invariance to choose a gauge in which the
spatial metric is diagonal (we focus on the scalar metric
fluctuations \footnote{See \cite{MFB} for a review of the theory of
cosmological perturbations and \cite{RHBrev1} for a shorter
overview.}). Making use of the perturbed constraint equations, we
determine the action for the remaining degrees of freedom of the
cosmological perturbations. At this stage there are indeed two
apparent physical degrees of freedom present. After expressing the
action in terms of the usual Sasaki-Mukhanov variable
\cite{Sasaki,Mukh2}, the variable in terms of which the action for
cosmological perturbations has canonical kinetic term, we find that
the extra degree of freedom for scalar metric fluctuations is
non-dynamical. In particular, there is no strong coupling problem
for the fluctuations: consistently including cosmological expansion
regulates the divergence found in \cite{Saffin}. In the case of
cosmological perturbations, the absence of new dynamical degrees of
freedom holds for any value of $\lambda$, whereas for gravitational
waves our analysis only covers the case $\lambda = 1$, the most
interesting case.

The action for cosmological perturbations derived in this paper allows
us to study the spectrum of curvature fluctuations in HL gravity. If
the potential term in the Ho\v{r}ava action is taken to be of
``detailed balance" form, we
find a cancellation of the leading UV terms in the action. Thus,
unlike what happens in the case of  HL matter on a fixed background,
the initial spectrum of fluctuations is not scale-independent. Scale
invariance of the fluctuations in the UV region is maintained if
we add terms which break the detailed balance condition.
More generally, we study ways of obtaining a scale-invariant spectrum
during the course of cosmological evolution.

The outline of this paper is as follows: We begin with a brief
review of HL gravity. In Section 3 we discuss cosmological
perturbations and show that no strong coupling problem arises.
In Section 4, we compute the power spectrum of cosmological
perturbations and discuss applications to inflationary cosmology.
The final section contains a discussion and conclusions.

\section{Setup}

\subsection{Brief review of Ho\v{r}ava-Lifshitz theory}

The dynamical degrees of freedom in HL gravity are the usual
metric degrees of freedom which appear in the ADM approach to
canonical gravity, namely the spatial metric $g_{ij}$, the lapse
function $N$ and the shift vector $N_i$. In terms of these fields, the full
space-time metric is
\be \label{metric}
ds^2 \, = \, - N^2 dt^2 + g_{ij} \bigl(dx^i + N^i dt \bigr)
\bigl( dx^j + N^j dt \bigr) \, ,
\ee
where the indices of $N_i$ are raised and lowered using the spatial
metric $g_{ij}$.

Note that here we will allow $N\equiv N(t,{\bf x})$, as the most
general form in the ADM decomposition. Because if one assumes
$N\equiv N(t)$, then the Hamiltonian constraint for perturbations
will be lost, and one can not recover GR in the IR limit and in the
case $\lambda=1$. For studies of cosmological perturbations assuming
$N=N(t)$, the reader is referred to \cite{Gao, Cai, Chen}.

The action of Horava-Lifshitz gravity contains a ``kinetic" part and
a ``potential" part,
\eq{
    S^g= S^g_{K} +S^g_{V} \,, }
with
    \eq{{\label{g_k_action}}
    S^g_K =  \frac{2}{\kappa^2} \int dt d^3x \sqrt{g} N
   \left( K_{ij}K^{ij}-\lambda K^2 \right)   \ ,
    }
where
    \[
    K_{ij}= \frac{1}{2N} \lrp{\dot{g}_{ij}- \nabla_i N_{j} - \nabla_j N_{i} } \, ,
    \]
is the extrinsic curvature and $K= g^{ij}K_{ij}$. Ho\v{r}ava chose
the potential to be of the ``detailed-balance" form \footnote{It is
argued in \cite{Nastase,Saffin} that solutions of GR are often not recovered if the
potential term is taken to be of detailed balance form, a problem already
encountered in \cite{Pope} in the context of spherically symmetric
metrics. This problem is due to a strong coupling signature which does not
arise if renormalizable terms which break the detailed balance condition
are added to the potential. Going beyond the detailed balance form of the
potential also allows IR solutions with a positive or vanishing cosmological constant
\cite{Visser,Kehagias}, whereas maintaining the detailed balance condition yields a
negative cosmological constant.}
    \ea{ {\label{Dbalance}}
    S^g_V & = \int dt d^3x \sqrt{g}  N  \lrsb{ -\frac{\kappa^2}{2 w^4}C_{ij}C^{ij}
     +\frac{\kappa^2\mu}{2 w^2}\epsilon^{ijk}R_{il} \nabla_j R^{l}_{k}
       -\frac{\kappa^2\mu^2}{8}R_{ij}R^{ij} +\frac{\kappa^2\mu^2}{8(1-3\lambda)}\left(\frac{1-4\lambda}{4}R^2+\Lambda R-3\Lambda ^2\right)
     }
     \,,
    }
where $C_{ij}$ is the Cotton tentsor defined by
    \eq{
        C^{ij}= \frac{\epsilon^{ikl}}{\sqrt{g}} \nabla_k \lrp{
        R^{j}_{l}-\frac{1}{4}R\delta^{j}_{l} }
        \,.
    }
Note that in (\ref{g_k_action}) $\lambda$ is a dimensionless
coupling of the theory and therefore runs as a function of energy. The
extra terms in the potential contain two further constants $w$ and $\mu$.
For our applications to cosmology, the terms involving $w$ will
not play a role. The action of GR is recovered in the IR limit if $\lambda = 1$.

The general structure of the action of scalar field matter in
Ho\v{r}ava-Lifshitz gravity contains two parts: a quadratic kinetic
term invariant under foliation-preserving diffeomorphisms and a potential
term:
    \eq{{\label{scalar_action}}
        S^{\varphi} = \int dt d^3x\, \sqrt{g}N
        \lrsb{ \frac{1}{2N^2} \lrp{ \dop - N^i \rd_i \varphi }^2  + F(\varphi,\rd_i\varphi,g_{ij})
        }\,,
    }
with the ``potential terms"
    \eq{{\label{scalar_f}}
        F (\varphi,\rd_i\varphi,g_{ij}) = -V(\varphi) + g_1 \xi_1 +
        g_{11} \xi_1^2 + g_{111} \xi_1^3 + g_2
        \xi_2 + g_{12} \xi_1 \xi_2 + g_3 \xi_3 \,,
    }
where the $\xi_i$ are invariants built out of spatial gradients of
$\varphi$: 
\bea
\xi_1 \, &=& \, \partial^i \varphi \partial_i \varphi \, , \\
\xi_2 \, &=& \, (\Delta \varphi)^2 \, , \\
\xi_3 \, &=& \, (\Delta \varphi) (\Delta^2 \varphi) \, . \eea 
Here and in what follows, we use $ \Delta \equiv \rd^i \rd_i \equiv
\rd^2/a^2 $ as a shorthand. Note that $g_1$ must be negative in
order to obtain the standard form of the kinetic term in the IR.

\subsection{Background equations of motion}

The equations of motion for $N$ and $N_i$ are the energy constraint
and momentum constraints, respectively. They take the general form
\ea{{\label{constraint}}
    0 &= -\frac{2}{\kappa^2} \lrp{ K_{ij}K^{ij} - \lambda K^2 } -\frac{\kappa^2}{2 w^4}C_{ij}C^{ij}
     +\frac{\kappa^2\mu}{2 w^2}\epsilon^{ijk}R_{il} \nabla_j R^{l}_{k}
       -\frac{\kappa^2\mu^2}{8}R_{ij}R^{ij} \\
       &\qquad\qquad + \frac{\kappa^2\mu^2}{8(1-3\lambda)}\left(\frac{1-4\lambda}{4}R^2+\Lambda R-3\Lambda
       ^2\right) -\frac{1}{2N^2} \lrp{ \dop -N^i\rd_i \varphi }^2 + F
       \,,\\
    0 &= \frac{4}{\kappa^2} \nabla_j\lrp{ K^j_i -\lambda K \delta^j_i } -
    \frac{1}{N} \lrp{ \dop - N^i \rd_i \varphi } \rd_i \varphi \,.
} 

In this work, we focus on a spatially flat background. Thus the
background values for the metric are 
    \eq{
        N=1\,,\qquad N_i =0\, , \qquad g_{ij} = a^2 \delta_{ij}\,,\qquad
        \varphi_0 = \varphi_0(t) \,,
    }
where $a=a(t)$ is the traditional scale-factor. In this background,
one has $C_{ij}=0$ and $\epsilon^{ijk}R_{il} \nabla_j R^{l}_{k} =0
$. Hence, at the background level, the energy constraint gives 
    \eq{{\label{e_constraint_0th}}
        0 = \frac{3 \kappa ^2 \Lambda ^2 \mu ^2}{8(3\lambda-1) }-V_0-\frac{\dot{\varphi }_0^2}{2}+\frac{6 (3 \lambda-1 ) H^2}{\kappa ^2
        }\,,
    }
while the momentum constraint is trivially satisfied.

The space diagonal component of the generalized Einstein equation
takes the form 
    \eq{{\label{FRW}}
        \frac{2(3\lambda-1)}{\kappa^2}\lrp{2\dot{H}+3H^2} + \frac{3\kappa^2 \Lambda^2
        \mu^2}{8(3\lambda-1)} +
        \frac{1}{2} \dop_0^2 - V_0 =0 \,,
    }
where $H\equiv \dot{a}/a$ is the Hubble parameter and $V_0 \equiv
V(\varphi_0)$.

Combining Eq. (\ref{FRW}) and the energy constraint
(\ref{e_constraint_0th}) equations, we find another useful equation:
    \eq{
        \frac{4(3\lambda-1)\dot{H}}{\kappa^2}  + \dop_0^2 = 0\,.
    }

The background equation of motion for the scalar field is as usual
    \eq{
        \ddot{\varphi}_0 + 3H \dop_0 +V' = 0 \,.
    }

\section{Perturbation Theory and the Dynamical Degrees of Freedom}

The scalar metric fluctuations about our background can be written as
(see \cite{MFB} for an overview of the theory of cosmological perturbations)
\bea
\delta g_{00} \, &=& \, - 2  \phi \, , \\
\delta g_{0i} \, &=& \, a^2 \partial_i B \, , \\
\delta g_{ij} \, &=& \, - 2 a^2 \bigl( \psi \delta_{ij} - \partial_i \partial_j E) \,
\eea
where $\phi, B, \psi$ and $E$ are functions of space and time.
Matter is perturbed, as well. The scalar field perturbation is
denoted by $\dep \equiv Q$.

The theory is invariant under spatial diffeomorphisms
\be
x^i \, \rightarrow \, x^i + f^i(x^j, t) \,
\ee
and under space-independent time reparametrizations. Compared to the
situation in GR, one has lost the invariance under space-dependent changes
in time.

We can use the spatial diffeomorphism invariance to choose the gauge
$E=0$, but one then no longer has the extra gauge freedom to choose
spatially flat gauge ($\psi = 0$ in addition to $E = 0$) or
longitudinal gauge ($B = 0$ in addition to $E = 0$). We are left
with three variables, namely $\phi$, $\psi$ and $B$. The lapse and
shift functions become
    \ea{
        N \, = \, 1 + \phi(t,x^i) \,,\\
        N_i \, = \, \rd_i B(t,x^i) \, ,
    }
and  that $\phi$, $B$ and $\psi$ are of the same order as the scalar
field perturbation $Q\equiv\dep$. As in GR, in order to get the
second order (and even the third order) action, we only need to
expand $N$ and $N_i$ to first order. Higher order contributions of
$N$ and $N_i$ can be eliminated making use of the equations of
motion.

One should also note that in the $E=0$ gauge, the spatial metric
$g_{ij}$ is conformal flat. In this gauge, making use of the local
Weyl transformation, one has $C_{ij}=0$ and $\epsilon^{ijk}R_{il}
\nabla_j R^{l}_{k} =0$. So the parameter $\omega$ does not enter the
cosmic perturbation theory.

\subsection{Constraints}

At first order, the energy constraint gives
    \ea{ {\label{e_constraint}}
        0 &=  2(1-3\lambda) \lrp{ \kappa^2 \dop_0^2 +12(1-3\lambda) H^2
        }\phi + 8(3\lambda-1)^2H \Delta B \\
        &\qquad\qquad +
        2\kappa^2(3\lambda-1) \lrp{ \dop_0 \doq + V'Q } +
        \kappa^4\mu^2\Lambda \frac{\rd^2\psi}{a^2} +24
        (3\lambda-1)^2 H \dot{\psi}
    }
and the momentum constraint gives
    \eq{{\label{m_constraint}}
        0 = (3\lambda-1)H \phi +
        (\lambda-1)\Delta B +(3\lambda-1)\dot{\psi} -
        \frac{\kappa^2}{4}\dop_0 Q \,.
    }

From (\ref{e_constraint}) and (\ref{m_constraint}) we can solve
for $\phi$ and $B$ explicitly to get
    \ea{{\label{constraint_alpha_beta}}
        \phi &= \frac{1 }{16 H^2 (1-3 \lambda )^2+2 \kappa ^2 (\lambda-1 ) (3 \lambda-1 ) \dot{\varphi }_0^2} \left\{  \Delta \psi  \kappa ^4 (-1+\lambda ) \Lambda  \mu ^2-16 H (1-3 \lambda )^2 \dot{\psi } \right. \\
        &\qquad\qquad \left. +2 \kappa ^2 (-1+\lambda ) (-1+3 \lambda ) \dot{Q} \dot{\varphi }_0+2 Q \kappa ^2 (-1+3 \lambda ) \left(H (-1+3 \lambda ) \dot{\varphi }_0+(-1+\lambda ) V'\right)  \right\} \,,\\
        \Delta B &= \frac{1}{32 H^2 (3 \lambda -1)+4 \kappa ^2 (\lambda -1) \dot{\varphi }_0^2} \left\{ -2 H \Delta \psi  \kappa ^4 \Lambda  \mu ^2-4 \kappa ^2 (-1+3 \lambda ) \dot{\varphi }_0 \left(H \dot{Q}+\dot{\psi } \dot{\varphi }_0\right) \right. \\
        &\qquad\qquad\qquad \left. +Q \kappa ^2 \left(12 H^2 (1-3 \lambda ) \dot{\varphi }_0+\kappa ^2 \dot{\varphi }_0^3+4 H (1-3 \lambda ) V'\right)  \right\} \,.
    }

Note that these constraint equations allow us to solve for $\phi$ and $B$ without any
singularities, even in the case $\lambda = 1$. If one (incorrectly) had expanded about
Minkowski space-time instead of about an expanding universe, one would have obtained
a singularity in the constraint equation (see e.g. Eq. (68) of \cite{Saffin}).
Note also that $\phi$ and $B$ remain perturbatively small. We learn the important lesson that
the expansion of space which is inevitable in the presence of matter removes the potential
strong coupling problem for cosmological perturbations. This is an important consistency
check for the cosmology of HL gravity.

\subsection{Second-order action}

At this stage, we have two independent degrees of freedom for scalar metric
fluctuations (instead of only one as would be the case in GR), namely $\psi$ and $Q$.
In order to set up the linear theory of cosmological perturbations, we need to
find the second order action for the fluctuations.

After inserting the perturbed metric and perturbed matter into the action for
gravity and matter, expanding to second order in the fluctuations, and make use
of the constraints (\ref{constraint_alpha_beta}), we obtain
    \ea{{\label{2nd_action_psi_Q}}
        S_2[\psi,Q] &= \int dtd^3x\, a^3 \left\{ c_{\varphi}\, \doq^2 + f_{\varphi}\, \doq Q -g_1 Q \Delta Q + g_2 \lrp{\Delta Q}^2 +g_3 (\Delta Q)(\Delta^2
        Q)   +  m_{\varphi}\,Q^2
        \right. \\
        &\qquad\qquad\qquad\qquad + c_{\psi}\,
        \dot{\psi}^2  + f_{\psi}\, \psi \dot{\psi} + h_{\psi}\,
        \dot{\psi} \Delta\psi  +
        \omega_{\psi}\, \psi \Delta\psi  + d_{\psi}\, (\Delta\psi)^2   +  m_{\psi}\, \psi^2  \\
        &\qquad\qquad\qquad\qquad \left. + c_{\psi\varphi}\, \doq \dot{\psi}  +
        f_{\psi\varphi}\, \psi \doq +
        \tilde{f}_{\psi\varphi}\, \dot{\psi}Q   + h_{\psi\varphi}\, \doq \Delta \psi   +
        \omega_{\varphi\psi}\, Q\Delta\psi + m_{\psi\varphi}\, \psi Q  \right\} \,.
    }
with $g_i = g_i(\varphi_0)$. The expressions for the various
coefficients can be found in Appendix \ref{app_coeff_1}.

It appears that there are two dynamical degrees of freedom, $\psi$ and
$Q$ respectively. However, it will be shown that this is an illusion
(see the next subsection for a detailed discussion on the issue of
dynamical degrees of freedom in Ho\v{r}ava theory). In fact, there
is only one dynamical degrees of freedom, the same as in
GR. The easiest way to see this is to look at the combination of
all of the kinetic terms in the above action, i.e. the
$\dot{Q}^2$, $\dot{\psi}^2$ and $\dot{Q}\dot{\psi}$ terms (see
Appendix \ref{app_coeff_1}), and to realize that they can
be brought into a ``perfect square" form
\[
    c_{\varphi}\, \doq^2 + c_{\psi}\,
        \dot{\psi}^2 + c_{\psi\varphi}\, \doq \dot{\psi} \quad \propto
        \quad
        \lrp{ \dot{\psi} + \frac{H}{\dop_0} \doq }^2 \, .
\]
This fact implies that there is indeed only one dynamical degrees of
freedom in our system. This degree of freedom is precisely the
Sasaki-Mukhanov combination of matter and metric fluctuations
\cite{Sasaki,Mukh2}, the variable in terms of which the action for
cosmological perturbations in GR has canonical form.

The Sasaki-Mukhanov variable $\zeta$
    \eq{{\label{def_zeta}}
        -\zeta \equiv \psi + \frac{H}{\dop_0}Q \,,
    }
is the gauge-invariant curvature perturbation on
uniform-density hypersurfaces. From (\ref{def_zeta}), we can express
$Q$ in terms of $\psi$ and $\zeta$,
    \ea{
        Q &= -\frac{\dop_0}{H}(\zeta+\psi) \,,\\
        \doq &= -\lrp{ \frac{\ddot{\varphi}_0 H - \dop_0 \dot{H}}{H^2}
        }(\zeta+\psi) - \frac{\dop_0}{H}\lrp{\dot{\zeta} +
        \dot{\psi}} \,,\qquad \textrm{etc.}
    }
After plugging the above relations into (\ref{2nd_action_psi_Q}),
and using the background equations of motion, we get a new action
for the two variables $(\zeta,\psi)$:
    \ea{{\label{2nd_action_zeta_psi}}
        S_2[\zeta,\psi] &= \int dtd^3x\, a^3 \left\{ c_{\zeta}\, \dot{\zeta}^2 +
        f_{\zeta}\,\dot{\zeta}\zeta  + \omega_{\zeta}\,\zeta\Delta\zeta  + d_{\zeta}\, (\Delta\zeta)^2  + \tilde{d}_{\zeta} \, \Delta\zeta\Delta^2\zeta  +
        m_{\zeta}\, \zeta^2
        \right. \\
        &\qquad\qquad\qquad + f_{\psi} \, \dot{\psi}\psi + h_{\psi}\, \dot{\psi} \Delta\psi  + \tilde{d}_{\psi}\, \Delta\psi \Delta^2\psi + d_{\psi}\,
        (\Delta\psi)^2   + \omega_{\psi}\,
        \psi \Delta\psi +  m_{\psi}\,
        \psi^2 \\
        &\qquad\qquad\qquad \left. +
        f_{\zeta\psi}\, \dot{\zeta}\psi  +
        \tilde{f}_{\zeta\psi} \, \zeta\dot{\psi} + h_{\zeta\psi}\, \dot{\zeta} \Delta\psi +
        \tilde{\omega}_{\zeta\psi}\, \zeta\Delta\psi  + \omega_{\psi\zeta}\, \psi
        \Delta\zeta    +d_{\zeta\psi} \Delta\zeta \Delta\psi+ \tilde{d}_{\zeta\psi} \, \Delta\zeta
        \Delta^2\psi + m_{\zeta\psi}\, \zeta\psi \right\} \,,
    }
where the various coefficients can be found in Appendix
\ref{app_coeff_2}.

It is important to note that there is no $\dot{\psi}^2$ term in
(\ref{2nd_action_zeta_psi}). Moreover, the coefficient of the
kinetic term for $\zeta$ is $c_{\zeta} \propto \dop_0^2$, and thus,
$\zeta$ also becomes non-dynamical in the absence of matter field
$\varphi$. In the presence of matter field $\varphi$, there is only
one dynamical degree of freedom in our system, which we can identify
as $\zeta$.

Eq. (\ref{2nd_action_zeta_psi}) is rather complicated. However, we
can further simplify the expression. First we note that all the
coefficients in (\ref{2nd_action_zeta_psi}) are evaluated in terms
of the background fields, and thus are functions of time only. Then,
for a general function $F(t)$ and spacetime field $\phi$, we have
the following convenient relation
\[
        \int dtd^3x\, a^3 F(t) \dot{\phi}\phi \simeq  \int dtd^3x\, \od{}{t}\lrp{ -\frac{a^3}{2}
        F(t) }
        \phi^2 = \int dtd^3x\, a^3 \lrsb{ -\frac{1}{2} \lrp{\dot{F} +3HF} }
        \phi^2 ~,
    \]
where ``$\simeq$'' denotes up to total derivative terms. Similarly,
we have (noticing that $\Delta \equiv \rd^2/a^2$)
    \[
        \int dtd^3x\,a^3 F(t) \dot{\phi}\Delta\phi \simeq \int
        dtd^3x\, a^3 \lrsb{ -\frac{1}{2} \lrp{\dot{F} +HF} } \phi
        \Delta\phi\, .
    \]
Thus, by using the above relations and performing many integrations by
parts, we find that the second-order action (\ref{2nd_action_zeta_psi}) can be
recast into a rather convenient form:
    \ea{{\label{2nd_action_after_ibp}}
        S_2[\zeta,\psi]
         &\equiv \int dt d^3x\, a^3 \left\{ c_{\zeta}\, \dot{\zeta}^2 + \zeta\, \Gamma_4(\Delta)\, \zeta
        +  \psi\, \Gamma_1(\Delta)\, \psi + \lrp{ \Gamma_2(\Delta)\zeta  +
        \Gamma_3(\Delta) \dot{\zeta}  } \psi \right\} \,,
    }
where we have defined
    \ea{{\label{Gamma}}
    \Gamma_1(\Delta) &\equiv   -\frac{1}{2}\lrp{ \dot{h}_{\psi}+Hh_{\psi} } \Delta  + \tilde{d}_{\psi}\, \Delta^3 + d_{\psi}\,
        \Delta^2    + \omega_{\psi}\,
        \Delta +  m_{\psi} -\frac{1}{2}\lrp{ \dot{f}_{\psi}  +3Hf_{\psi}
        }  \,,\\
    \Gamma_2(\Delta) &\equiv \lrp{ \omega_{\psi\zeta} +\tilde{\omega}_{\zeta\psi} }
        \Delta + d_{\zeta\psi} \Delta^2 + \tilde{d}_{\zeta\psi} \, \Delta^3
         + \lrp{ m_{\zeta\psi} -\dot{\tilde{f}}_{\zeta\psi} -3H \tilde{f}_{\zeta\psi} }
         \,,\\
    \Gamma_3(\Delta) &\equiv  h_{\zeta\psi}\,
        \Delta \,,\\
    \Gamma_4(\Delta) &\equiv \omega_{\zeta}\,\Delta  + d_{\zeta}\, \Delta^2  + \tilde{d}_{\zeta} \, \Delta^3  +
        m_{\zeta}  -\frac{1}{2} \lrp{\dot{f}_{\zeta} +3H f_{\zeta}}
        \,,
    }
for simplicity (in $\Gamma_3$, we have used the fact $f_{\zeta\psi}
=\tilde{f}_{\zeta\psi}$, see Appendix \ref{app_coeff_2} for
details). Note that since $\Delta \equiv \rd^2/a^2$ and since
several of the coefficients in (\ref{2nd_action_zeta_psi}) are time-dependent,
the $\Gamma$'s are in general time-dependent. The above results should
be understood in Fourier space, where we identify $\Delta \equiv
-k^2/a^2$.

The important point is that now $\psi$ has no time-derivatives and
 acts as a new constraint. The equation of motion for $\psi$ is
    \eq{{\label{eom_psi_constraint}}
         2\, \Gamma_1(\Delta) \psi   +
          \Gamma_2(\Delta)\zeta  +
        \Gamma_3(\Delta) \dot{\zeta} =0 \,,
    }
from which  we can solve $\psi$ explicitly to get
    \eq{{\label{psi_solution}}
        \psi =  - \frac{
         \Gamma_2(\Delta)\zeta  +
        \Gamma_3(\Delta) \dot{\zeta}  }{ 2\,\Gamma_1(\Delta)  }
        \,.
    }

After plugging (\ref{psi_solution}) into
(\ref{2nd_action_after_ibp}), and after some straightforward
calculations, we obtain an effective quadratic action for a single
variable $\zeta$,
    \ea{{\label{2nd_action_zeta_only}}
        S_2[\zeta] &= \int dt d^3x\, a^3 \lrcb{ \lrp{c_{\zeta} - \frac{\Gamma_3^2}{4\Gamma_1} } \dot{\zeta}^2 + \lrsb{ \Gamma_4 - \frac{\Gamma_2^2}{4\Gamma_1} + \frac{1}{4a^3}  \od{}{t} \lrp{\frac{a^3 \Gamma_2\Gamma_3}{\Gamma_1}}     }
        \zeta^2
        } \,.
    }

\subsection{Subtleties of the dynamical degrees of freedom}

The important lesson to be drawn from the previous subsection is
that the extra degree of freedom which appears in HL gravity due to
the loss of space-dependent time reparametrizations as a symmetry of
the theory is \emph{non-dynamical} in cosmology. Hence, the strong
coupling problem discussed recently in \cite{Saffin} is not present
in linear cosmological perturbation theory. Let us discuss this
result, a result which holds for any value of $\lambda$.

The symmetry group of Ho\v{r}ava-Lifshitz gravity is that of
 ``foliation-preserving" diffeomorphisms, which
is smaller than the full diffeomorphism group of GR. Thus, one may
naively expect that with smaller symmetry, we are left with less
gauge artifacts and more physical modes. In particular, it was
argued that there was an additional scalar dynamical degree of
freedom in Ho\v{r}ava theory, arising from the gravity sector
itself. This sounds very different from the situation in GR, where
gravity has only two physical degrees from freedom --- the two
polarization states of gravitational waves. According to the
standard treatment of cosmological perturbation theory in GR, the
quantum perturbations of the scalar sector are expected to be
generated by scalar matter fields. In this case, if there is no
matter, all perturbation modes are gauge artifacts. Thus, if there
is indeed one additional scalar dynamical degree of freedom from
gravity itself in Ho\v{r}ava theory, our traditional picture of
cosmological perturbation would not apply. In fact, the extra
dynamical degree of freedom would lead to serious problems for the
theory, as stressed in \cite{Saffin}. However, in this work, a
detailed study of the perturbation theory shows that there is no
additional dynamical scalar degree of freedom at all, when
perturbations in the presence of matter are found by consistently
expanding around the FRW metric and not around the flat Minkoswki
one.

Furthermore, another point is that, in Ho\v{r}ava's original
formulation of the theory, the lapse function $N$ was restricted to
being a function of time only $N=N(t)$, where the corresponding
Hamiltonian becomes non-local. In this work, we relax this
 restriction. By assuming $N=N(t,\vx)$ and starting from the most general expansion of the
action, we see that the apparent additional degree of freedom is
non-dynamical.

Is there a new dynamical degree of freedom in the absence of
matter? Let us take the limit of our equations when $H$ and
the energy density of matter tend to zero.
Since the coefficient of the kinetic term $\dot{\zeta}^2$ is proportional
to the background scalar-field value $\dop_0$ (see Appendix
\ref{app_coeff_2}), $c_{\zeta} \propto \dop_0^2$. Thus if there is
no matter field, all perturbation modes become non-dynamical, which
is exactly the case in GR. However, the limit we are discussing is
singular unless $\lambda = 1$ because if $\lambda \neq 1$ one of the
coefficients in the action (the $d_{\psi}$ coefficient) blows up.

Thus, our work also shows that in the case $\lambda = 1$, the case
in which the action of HL gravity reduces to that of GR in the IR
limit, there are no extra gravitational degrees of freedom in the
vacuum. This result is in agreement with the conclusions in
\cite{Cai}, where a scalar degree of freedom in the gravity sector
was identified for $\lambda \neq 1$, and where it was shown that
this mode becomes non-dynamical when $\lambda = 1$. Our conclusions
also agree with the recent analysis of \cite{Visser2} \footnote{The absence
of an extra dynamical degree of freedom in the vacuum sector of HL gravity
is also observed in \cite{Kehagias,Myung}. }. The extra
scalar gravitational degree of freedom in the vacuum sector of the
theory was also discussed in \cite{Chen}. That paper shows that
there is a singularity for $\lambda = 1$ which is agreement with our
conclusion that for this choice of $\lambda$ there are no new
gravitational degrees of freedom.

\section{Cosmological Perturbations and The Power Spectrum}

In \cite{Mukh} (see also \cite{Piao}) it was pointed out that a
scalar field with Ho\v{r}ava-Lifshitz form (\ref{scalar_action}) will obtain a
scale-invariant spectrum of cosmological fluctuations, and it
was then argued that a scale-invariant spectrum of curvature
fluctuations may similarly result independent of the equation of
state of the background \footnote{The possibility of obtaining a
scale-invariant spectrum of metric perturbations in pure HL
gravity was discussed in \cite{Chen3}.}. This result follows from the fact that
the action contains terms with six space derivatives. These terms
dominate in the UV and yield a scale-invariant spectrum.

In cosmology we are interested in the power spectrum of the induced
curvature fluctuations. The formalism we established in the previous
section now allows us to calculate this spectrum. We will show that
for the gravitational potential (\ref{Dbalance}) of detailed balance
form, the terms with six spatial derivatives cancel in the action
for cosmological fluctuations. This happens independently of the
detailed form of the coefficients of the higher derivative terms in
the matter action. The reason for this result is that the terms in
the gravitational action with six spatial derivatives (the terms
involving the constant $w$) do not enter the action for cosmological
perturbations. If we add terms to the action consistent with power
counting renormalizability which break the detailed balance
condition, terms with six spatial derivatives in the action for
cosmological perturbations will survive.

If we keep the detailed balance condition on the potential, then
the leading terms in the UV contain four spatial derivatives. Thus,
the initial power spectrum of curvature fluctuations is not
scale-invariant \footnote{As already mentioned in \cite{Mukh}, the
scale-invariance of an initial matter entropy field spectrum can
induce scale-invariance of the curvature fluctuations via the
``curvaton" mechanism \cite{curvaton}.}. Thus, scale-invariance
of the late time power spectrum will only arise for specific
background evolutions. e.g. for inflationary expansion (as discussed
later in this section) or for a matter bounce \cite{RHB}
background. If we drop the detailed balance condition, then
an initially scale-invariant curvature power spectrum results.

\subsection{Equation of motion}

 The action (\ref{2nd_action_zeta_only}) for a gravitational
 potential satisfying the detailed balance condition
 has the general structure:
    \eq{{\label{general_stru}}
        S_2[\zeta] = \int dt d^3x\, a^3 \lrp{ \gamma\, \dot{\zeta}^2 - \Omega\,
        \zeta^2
        } \,,
    }
with
    \ea{{\label{gamma_Omega}}
        \gamma &\equiv \lrp{c_{\zeta} - \frac{\Gamma_3^2}{4\Gamma_1} } \,,\\
        - \Omega &\equiv \Gamma_4 - \frac{\Gamma_2^2}{4\Gamma_1} + \frac{1}{4a^3}  \od{}{t} \lrp{\frac{a^3
        \Gamma_2\Gamma_3}{\Gamma_1}} \,,
    }
where the $\Gamma$'s are defined in (\ref{Gamma}). In order to write the
action (\ref{general_stru}) in canonical form, we introduce the new
variable
    \eq{
        u \equiv a \sqrt{\gamma} \zeta \,.
    }
After changing to conformal time $\eta$ (which is defined by $dt = a
d\eta$) we have
    \ea{
        S_2[\zeta] &= \int d\eta d^3x\, \lrcb{ u'^2 + \lrsb{ \lrp{ \mathcal{H} + \frac{\gamma'}{2\gamma} }^2
          + \lrp{ \mathcal{H} + \frac{\gamma'}{2\gamma} }'
        - \frac{a^2 \Omega}{\gamma} } u^2  } \,,
    }
where $\mathcal{H} \equiv a'/a$, and a prime indicates the derivative with
respect to conformal time. Note that the above result
should be understood in momentum space, that is we must make the replacement
$\Gamma_i(\Delta) \rightarrow \Gamma_i(-k^2/a^2)$.

The classical equation of motion for the canonically-normalized variable
$u$ is simply
    \eq{{\label{master_equation}}
         u_k'' + \omega^2(\eta,k)\,  u_k =0 \,,
    }
with
    \eq{{\label{omega}}
        \omega^2(\eta,k) \equiv  \frac{a^2 \Omega}{ \gamma} -\lrp{ \mathcal{H} + \frac{\gamma'}{2\gamma} }^2
          - \lrp{ \mathcal{H} + \frac{\gamma'}{2\gamma} }'  \,,
    }
where $\gamma$ and $\Omega$ are defined in (\ref{gamma_Omega}) and
the variously introduced parameters can be found in Appendix
\ref{app_coeff_2}. We emphasize that in deriving the equation of
motion (\ref{master_equation})-(\ref{omega}), no approximation
(beyond the restriction to linear perturbation theory) has been made
and that thus the equation is exact. One can use
(\ref{master_equation})-(\ref{omega}) as the starting point of a
detailed investigation of the spectrum of scalar metric
perturbations in Ho\v{r}ava-Lifshitz theory.

In the following, we shall first investigate the UV limit of the
above equations in a general background, and then study the
evolution in an inflationary background.

\subsection{UV Limit of Perturbation Theory}

To study the perturbations in the UV limit, i.e. for
$k\rightarrow \infty$, we consider the terms with highest
power of $\Delta$ in the equation of motion for the
fluctuations. First, we note that the leading terms in
the $\Gamma$'s are
\begin{align}
  \Gamma_1\simeq \tilde{d}_{\psi}\Delta^3~,\quad
  \Gamma_2\simeq\tilde{d}_{\zeta\psi}\Delta^3~,\quad,
  \Gamma_3\simeq h_{\zeta\psi}\Delta~,\quad
  \Gamma_4\simeq \tilde{d}_\zeta \Delta^3~,\quad \gamma\simeq c_\zeta~,
\quad \Omega\simeq \left( -\tilde{d}_\zeta+\frac{\tilde{d}_{\zeta\psi}^2}{4d_\psi} \right)\Delta^3 \, ,
\end{align}
where the $\simeq$ sign means that we are writing down only the
leading term in the UV limit (but with the exact value of its
coefficient).

Inserting the coefficients (see Appendix \ref{app_coeff_2}):
\begin{align}
  \tilde{d}_\zeta\equiv\frac{g_3\dot\varphi_0^2}{H^2}~,\quad
  \tilde{d}_{\zeta\psi}\equiv\frac{2 g_3\dot\varphi_0^2}{H^2}~,\quad
  \tilde{d}_\psi\equiv\frac{g_3\dot\varphi_0^2}{H^2},
\end{align}
we find that
\be
\Omega \, \simeq \, 0 \, .
\ee
In other words, the $k^6$ term in $\Omega$, which could be
naively expected to arise in  Ho\v{r}ava-Lifshitz theory based on the
terms in the Lagrangian, vanishes. exactly. The leading order contribution
in powers of $k$ starts with a $k^4$ term. As shown in \cite{Mukh}, it
is the $k^6$ term which can naturally produce a scale invariant spectrum,
while the $k^4$ term cannot. Thus, for a potential satisfiyng the
detailed balance condition it is not true that in
Ho\v{r}ava-Lifshitz theory a scale invariant spectrum can be produced
in any background.

The above cancellation of the $k^6$ term is not an accident. One can
prove that even for more general high order derivative terms in the
scalar field Lagrangian, the above cancellation happens. To show that,
we take the high order derivative part of the Lagrangian to be
\footnote{See also \cite{Chen2} for a study of scalar field Lifshitz actions.}
\begin{align}
  \sum_{m=1,n>m}g_{mn}\Delta^m \varphi\Delta^n \varphi ~.
\end{align}
The corresponding piece of the matter Lagrangian quadratic in $Q$ is
\begin{align}\label{generalcancel}
  \sum_{m=1,n>m}g_{mn}\Delta^m Q \Delta^n Q
~\rightarrow
\sum_{m=1,n>m}\frac{g_{mn}\dot\varphi^2}{H^2}\Delta^{m+n}
(\zeta+\psi)^2~.
\end{align}
Note that here $\psi$ is not a dynamical field, and needs to be
solved for by minimizing the action. In the $k\rightarrow \infty$
limit, there are no other $\Delta^3$ or higher spatial derivative
terms, because the highest derivative in the gravity sector is
$\Delta^2$. Thus, Eq. \eqref{generalcancel} corresponds to the
dominant term in the action. Minimizing the action, the solution is
\begin{align}
  \psi \, \simeq \, -\zeta~.
\end{align}
Thus, the $\Delta^3$ or higher spatial derivative terms in the
scalar field sector are canceled and thus do not contribute to
cosmic perturbations.

If one assumes that there are terms such as $\varphi\Delta^3\varphi$
instead of $\Delta\varphi\Delta^2\varphi$ in the scalar field
Lagrangian, then the above proof is no longer valid. It is then
possible to generate scale invariant perturbations for a general
background in this case. However, such a scalar field Lagrangian
breaks the shift symmetry in the gradient term, and is thus not
conventionally used in the literature.

At this stage, it is also important to recall that $k^6$ terms in the action
for cosmological perturbations will survive if one adds terms to the
gravitational action which do not conform to the detailed balance
condition.

\subsection{Perturbation Theory in an Inflationary Background}

Now we turn to the perturbation equation of motion
(\ref{master_equation}). Obviously, (\ref{master_equation}) reduces
to the standard case in IR. Thus, in this work, in order to
investigate the possible differences of cosmological perturbation in
Ho\v{r}ava theory from those in General Relativity, we focus on the
UV region. For simplicity, we consider an inflationary background as
usual, where the Hubble parameter is approximately constant.

As discussed in the previous subsection, in the UV limit, the $\sim
k^6$ term \emph{exactly} cancels out. The equation of motion takes
the form
    \eq{{\label{eom_UV}}
        u''_k + \lrp{ c_s^2 k^2 + \Xi^2\, \frac{k^4}{a^2} - \frac{a''}{a} + M^2 a^2 } u_k = 0 \,.
    }
where
    \ea{{\label{Xi}}
        \Xi^2 &\equiv \frac{(\lambda -1) \kappa ^2  \mu ^2 \left[16 H^2 (1-3 \lambda )^2+\kappa ^4 \Lambda ^2 \mu ^2+2 \kappa ^2 (\lambda -1) (3 \lambda-1 ) \dot{\varphi }_0^2 \right] }{32 (3 \lambda-1 )^3 \dot{\varphi
        }_0^2} \,,
    }
$c_s^2$ is the analogue of the ``speed of sound" and $M^2$ is
the effective mass-square term. The expressions for both $c_s^2$ and $M^2$
can be found in Appendix \ref{app_mass}. We should keep in mind that (\ref{eom_UV}) only
describes the behavior of perturbation in UV limit, that is, the
left-hand-side of (\ref{eom_UV}) should be compensated with terms of
order $\mathcal{O}(1/k^2)$, $\mathcal{O}(1/k^4)$, $\cdots$ etc.,
which we neglect in the following analysis.

It is interesting to point out a connection with the ``trans-Planckian problem" for
fluctuations in inflationary cosmology. In  \cite{RHBrev0}, it was argued that
the predictions of inflationary cosmology for the spectrum of cosmological
perturbations are sensitive to hidden assumptions about physics on trans-Planckian
scales. To demonstrate this point, fluctuations obeying a modified dispersion relation
very much like the one we have obtained above was studied in \cite{Jerome}.
However, in \cite{Jerome}, the modified dispersion relation was not derived from
any action principle but simply postulated. We have shown here that HL gravity
yields a specific modified dispersion relation for scalar metric fluctuations.
We will return to this point in a followup paper \cite{Gao2}.

Note that in the case $\lambda=1$, the case in which the IR limit of
HL gravity yields the Einstein action, $\Xi=0$ and thus the $\sim
k^4$ term in the equation of motion for the curvature fluctuations
also vanishes. Thus, in this case, the case of most practical
interest, the equation of motion for the curvature fluctuations
reduces to the same form as in GR. The only difference is that the
effective mass $M$ takes a different from.

In the case $\lambda \neq 1$, then in order to guarantee the stability of the perturbations in
UV limit, $\Xi^2$ has to be positive. If we assume
$1/3  <  \lambda < 1$, from (\ref{Xi}), it is easy to show this gives a ``lower bound"
for $|\dop|$:
    \eq{\label{varphilowerbound}
        \frac{\dop^2_0}{H^2} > \frac{16H^2 (3\lambda-1)^2 + \kappa^4 \Lambda^2\mu^2 }{2H^2 \kappa^2 (1-\lambda)
        (3\lambda-1)} \,,
    }
or, in terms of $V_0$:
    \eq{
        16 H^2 (1-3 \lambda )^2+\kappa ^4 \Lambda ^2 \mu ^2-8 \kappa ^2 (-1+\lambda )
        V_0 < 0 \,.
    }
In the case $\lambda = 1$, the stability condition on the solutions in the UV requires that
the coefficient $g_1$ in the matter Lagrangian be negative, the sign we expect (since
it is the $g_1$ term which dominates in the IR and must give the standard kinetic term
for the scalar field Lagrangian).

As an application of the framework we have developed in this paper,
let us consider the evolution of fluctuations in an inflationary
background\footnote{We neglect slow-roll corrections to the scale
factor evolution.} with $a(\eta)=-1/(H\eta)$, where $\eta$ is the
conformal time. The equation of motion (\ref{eom_UV}) has been
extensively investigated. On super-Hubble scales the solution is 
\be
u_k(\eta) \, \sim \, a(\eta)
\ee
which corresponds to constant curvature fluctuation $\zeta$. In the special case
$\lambda = 1$, the solution $u_k(\eta)$ will be oscillating on sub-Hubble scales.

More generally, the equation can be solved analytically to
get{\footnote{Equation of motion with the same structure of
(\ref{eom_UV}) has also been analyzed in \cite{Gao:2009vi}, in the
investigation the statistical anisotropy and large-scale CMB
anomalies. }} 
    \eq{{\label{mode_function}}
        u_k(\eta) = \frac{1}{\sqrt{C}}\, e^{-z/2} \sqrt{-\eta} \,(-c_s k \eta)^{\nu}\, U(\alpha,
        \nu+1,z) \,,
    }
where $U(\alpha,\nu+1,z)$ is a
confluent hypergeometric function of the first kind with
    \ea{
        \alpha &\equiv \frac{1}{2}(\nu +1)-\frac{i\, c_s^2}{4 H \Xi
        } \,,\\
        \nu &\equiv
        \sqrt{\frac{9}{4}-\frac{M^2}{H^2}} \,,\\
        z &\equiv -i H \,\Xi\,  k^2 \eta^2 \,,
    }
and the overall normalization constant is
    \eq{
        C \equiv 2\, \nu\, e^{-\frac{i}{2}\pi(7\nu+1) } \lrp{ -\frac{c_s^2}{H\Xi}
        }^{\nu} \Gamma(\nu)\Gamma(-\nu) \lrsb{ \frac{1}{\Gamma(\alpha-\nu)\Gamma(\alpha^{\ast})} - \frac{e^{5i\pi\nu}}{\Gamma(\alpha) \Gamma(\alpha^{\ast} - \nu)}
        } \,,
    }
where ``$\ast$" denotes a complex conjugate. Here the $k$-independent
normalization constant $C$ is chosen so that $u_k(\eta)$ is
normalized in the usual way (unit Wronskian):
    \eq{
        u_k(\eta) u'^{\ast}_k(\eta) - u^{\ast}_k(\eta) u'_k(\eta) = i \,,
    }
which is the condition for canonical quantization.

\begin{figure}[h]
\centering
    \begin{minipage}{0.7\textwidth}
    \centering
    \includegraphics[width=8cm]{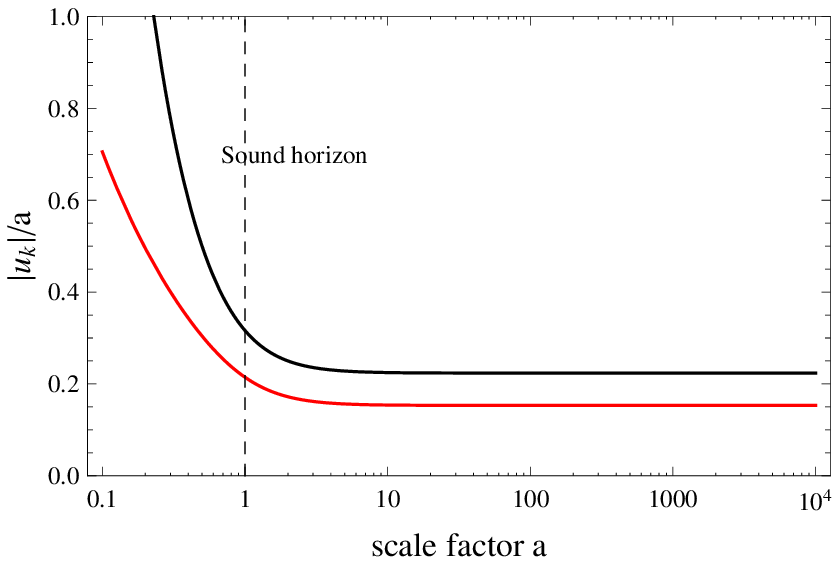}
    \caption{Evolution of the mode function with the scale factor $a$. The red curve
    corresponds to the mode function in Ho\v{r}ava theory, the black curve shows the
     standard mode function in GR. The dashed vertical line denotes the sound horizon.
     The parameters were chosen to be $c_s=1$, $k=H=10$, $M=0$, $\Xi=0.1$.}
    \label{fig_evolution}
    \end{minipage}
    \begin{minipage}{0.7\textwidth}
    \centering
    \includegraphics[width=8cm]{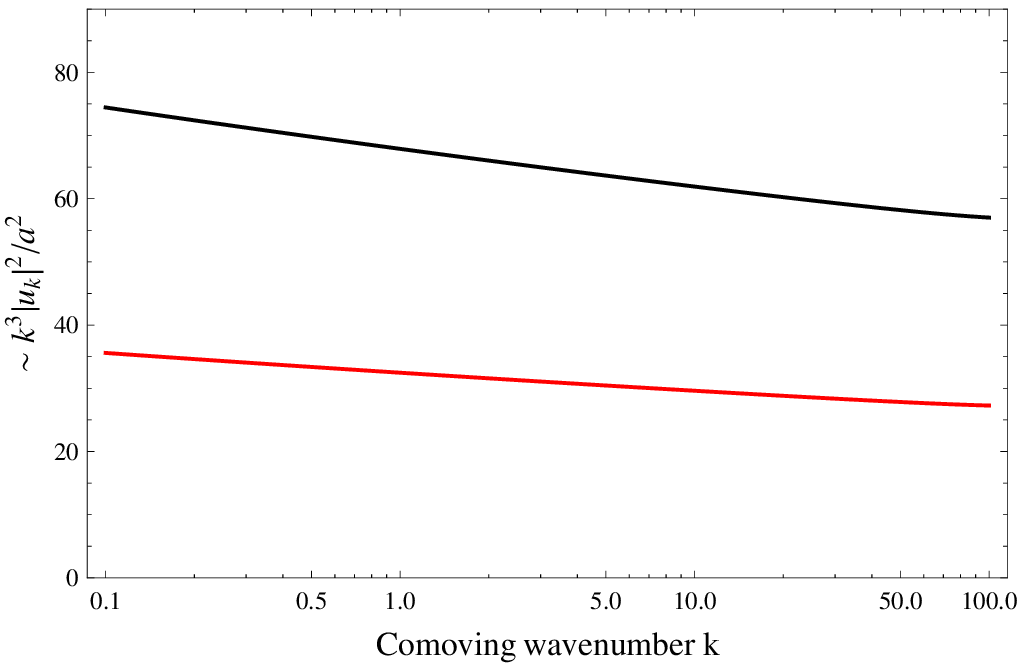}
    \caption{Power spectrum on super-Hubble scales. The red line denotes the
    spectrum of perturbations in Ho\v{r}ava theory, and the black line denotes the standard
    GR result. The parameters were chosen to be
    $c_s=1$, $\Xi=0.1$, $\nu=1.52$, $H=10$. The spectra are evaluated at the
    conformal time $\eta= -0.001$.}
    \label{fig_spectrum}
    \end{minipage}
\end{figure}

The evolution of a perturbation mode is shown in
Fig.\ref{fig_evolution}. It can be seen that the behavior of
perturbation in Ho\v{r}ava theory is very similar to that in GR.
In particular, after the wavelength exits the sound horizon, the perturbation
modes are frozen. On super-horizon scales, since
    \eq{
         u_k(\eta) \xrightarrow[]{\eta \rightarrow 0}  \frac{\Gamma(\nu) \sqrt{-\eta} }{\sqrt{C}\, \Gamma(\alpha)} \lrp{ \frac{c_s}{ iH \Xi k \eta } }^{\nu}  \,,
    }
the dimensionless power spectrum $P_{\zeta}$ of the curvature fluctuation
$\zeta$ can easily be calculated:
    \ea{
        P_{\zeta}(k) &\equiv \frac{k^3}{2\pi^2} |\zeta(k)|^2 \\
        &= \frac{k^3}{2\pi^2} \lrmb{ \frac{u_k(\eta) }{a \sqrt{\gamma}} }^2 \\
        &\approx \lrp{ \frac{H}{2\pi} }^2 \frac{ 2 \Gamma^2(\nu) }{ c_s^3 c_{\zeta} \lrmb{C}  \Gamma(\alpha) \Gamma(\alpha^{\ast})
        } \lrp{ \frac{c_s^2 }{ H\Xi} }^{2 \nu} \lrp{ - c_s k\eta}^{3-2
        \nu} \,,
    }
where we have used the fact that in UV limit, $\gamma \approx
c_{\zeta}$. The dimensionless power spectrum $P_{\zeta}(k)$
is shown in Fig.\ref{fig_spectrum} (up to an overall factor). The
figure also shows for comparison the power spectrum
obtained with the standard GR mode function
\be
u_k(\eta) = \frac{\sqrt{\pi}}{2} e^{i \lrp{ \nu+\frac{1}{2}}
\frac{\pi}{2}} \sqrt{-\eta} H_{\nu}^{(1)}(-c_s k\eta)~,
\ee
which describes the fluctuation of a massive light scalar field in
de Sitter space-time. We observe that the perturbations are
suppressed in Ho\v{r}ava-Lifshitz gravity compared with those in GR.
This verifies the argument that renormalizability of gravity
generally reduces the amplitude of perturbations \cite{Wang:2008cs}.
It would be also interesting to see whether this suppression of
perturbations could stabilize the inflationary background in the UV
limit and prohibit eternal inflation.

We also would like to mention that we have not proved whether
$\zeta$ is a conserved quantity beyond the slow roll approximation here.
Although in the solution, we can see $\zeta$ is no more than slowly
varying on super Hubble scales. In one Hubble time, the
variation of $\zeta$ is at most as large as the slow roll
parameters. It is still possible for $\zeta$ to receive ${\cal
O}(1)$ corrections during about 60 e-folds of inflation. Another
possible contribution to $\zeta$ on super Hubble scales is the UV-IR
transition. As the UV-IR transition happens at ${\cal O}(1)$
e-folds, it may also yield a ${\cal O}(1)$ contribution to
$\zeta$.

Applications of the equation of motion for cosmological fluctuations in HL
gravity to non-inflationary backgrounds will be left to a followup paper \cite{Gao2}.

\subsection{IR Limit}

Now we turn to the IR limit behavior of perturbations as an
important consistency check of perturbation theory in Ho\v{r}ava
gravity. In the IR limit ($k \rightarrow 0$),
(\ref{master_equation})-(\ref{omega}) takes the form
    \eq{{\label{eom_IR}}
        u''_k + \lrp{ \tilde{c}_s^2 k^2 - \frac{a''}{a} + \tilde{M}^2 a^2
        }u_k =0  \,,
    }
where the expressions for $\tilde{c}_s^2$ and $\tilde{M}^2$ can be
found in Appendix \ref{appsec_cs_mass_IR}. There are additional terms
on the left-hand-side of the equation which are of the order
$\mathcal{O}(k^4)$, $\mathcal{O}(k^6)$, etc. which we can
neglect in IR limit. Then, Eq. (\ref{eom_IR}) has the same form as the
corresponding equation
in standard perturbation theory in GR.

As we know, in the IR limit, Ho\v{r}ava theory reduces to GR with the
following parameters:
    \eq{
        c = \frac{\kappa^2 \mu}{4} \sqrt{ \frac{\Lambda}{1-3\lambda}
        } \,,\qquad 16\pi G = \frac{\kappa^4 \mu}{8} \sqrt{ \frac{\Lambda}{1-3\lambda}
        } \,, \qquad \Lambda_{\textrm{GR}} = \frac{3 \kappa^4 \mu^2
        \Lambda^2}{32(1-3\lambda)} \,.
    }
Thus, setting the speed of light $c$ in the IR to $c=1$ corresponds to choosing
$\Lambda = \frac{16 (1-3\lambda)}{\kappa^2 \mu^2}$ in Ho\v{r}ava
theory. With this value of $\Lambda$, from (\ref{app_tilde_cs}) it
is easy to show that
    \eq{
        \tilde{c}_s^2 \equiv -2g_1 \,,
    }
which is what we expected. Thus, the equation of motion for
perturbations in Ho\v{r}ava-Lifshitz theory in the IR limit indeed
reduces to the same form as in GR. However, the effective mass
$\tilde M$ and the parameter $\gamma$ still remains different from
GR. One can check that when one further takes $\lambda=1$, we have
 $\gamma=\frac{\dot\varphi_0^2}{2H^2}$, and $\tilde M$ is suppressed
 by slow roll parameters. So in the $g=-1/2$ case in the IR limit, the
perturbation theory completely returns to GR up to the leading order
of slow roll approximation. This is a consistency check for our
calculation.

To summarize, we have found several cases in which the perturbation
equation takes the same form as that in GR: (1) In the UV limit with
$\lambda=1$. In this case, the equation of motion has the same form
with that in GR. However, the coefficients are different. (2) In the
IR limit with arbitrary $\lambda$. In this case, the equation of
motion is the same with that in GR up to leading order in slow roll
approximation. (3) In the IR limit with $\lambda=1$. In this case,
the equation of motion should completely reduced to GR.

\section{Conclusions and Discussion}

In this paper we have studied the theory of linearized cosmological perturbations
in Ho\v{r}ava-Lifshitz (HL) gravity. In this study, it is important to expand about
a dynamical background since the presence of matter implies that the average
energy density does not vanish. We have found that the extra degree of freedom
which could be expected to arise because of the reduced symmetry of HL
gravity is in fact not dynamical. This conclusion holds for any value of $\lambda$.
Taking the flat space-time limit of our analysis, we can also show the
absence of any new dynamical scalar metric degree of freedom in the absence
of matter. Our limiting procedure, however, only works in the special case
$\lambda = 1$, the case in which the HL gravitational action flows to the action
of General Relativity (GR) in the infrared. We thus do not see any evidence of
the ``strong coupling problem" mentioned in \cite{Saffin}.

Starting from the most general metric including scalar cosmological perturbations,
we have worked out the quadratic action for cosmological perturbations.
It turns out that the distinguished dynamical variable for fluctuations is the
usual Sasaki-Mukhanov variable. In terms of this variable, the kinetic terms
in the action has canonical form. The second metric variable enters the
action without kinetic term and is hence not a dynamical degree of freedom.

Based on the action for cosmological perturbations, we can compute the
spectrum of these fluctuations. Rather surprisingly it turns out that
the terms in the equation of motion for these fluctuations containing six
spatial derivatives vanish if the potential term in the HL gravitational action
is of detailed balance form. It is the presence of these terms which leads
to the scale-invariance of the spectrum of HL matter \cite{Mukh,Piao}.
Thus, we conclude that the spectrum of cosmological perturbations is
not scale-invariant in HL cosmology (with potential satisfying the
detailed balance form), unlike the spectrum of spectator
matter field fluctuations. However, the fact that spectator HL matter
fields acquire a scale-invariant spectrum will likely make it possible
to use the curvaton mechanism \cite{curvaton} to induce scale-invariant
fluctuations in HL cosmology independent of the expansion rate
of space.

If the gravitational action is not of detailed balance form (and this
seems to be the preferred case \cite{Nastase,Saffin} if the IR limit
of the theory is really to reproduce Einstein gravity), then the $k^6$
terms in the action for cosmological fluctuations will persist, making
it possible to have a scale-invariant initial spectrum of adiabatic
cosmological perturbations along the lines suggested by \cite{Mukh}.

We would like to warn the reader that Ho\v{r}ava-Lifshitz gravity
faces many challenges before it can be declared as a viable candidate
theory for quantum gravity (see \cite{Saffin,Miao} for some potential
problems). We have only addressed one of these problems - the
strong coupling problem for additional fluctuation modes,
a problem which has been considered fatal for the theory - and shown
that in fact does not arise. Other potential problems remain to be
resolved. If they can be successfully resolved, then it becomes
of great interest to explore applications of the equations of cosmological
perturbations which we have derived here to non-inflationary
backgrounds. Work on this issue is in progress \cite{Gao2}.

\acknowledgements

We would like to thank L. Alvarez-Gaume, G. Dvali, E. Witten and K.
Zarembo for useful discussions. R.B. thanks the CERN Theory Division
for financial support and hospitality in a wonderfully stimulating
atmosphere. He also thanks C. Charmousis and A. Padilla for
correspondence. XG and YW are supported in part by a NSFC grant No.
10535060/A050207, a NSFC group grant No. 10821504, and a 973 project
grant No. 2007CB815401. The research of R.B. is supported in part by
a NSERC Discovery Grant and by funds from the Canada Research Chair
program.


\appendix

\section{Various Coefficients}

\subsection{Coefficients in (\ref{2nd_action_psi_Q})}\label{app_coeff_1}

    \eq{
        c_{\varphi} \equiv \frac{4 H^2 (-1+3 \lambda ) }{8 H^2 (-1+3 \lambda )+\kappa ^2 (-1+\lambda ) \dot{\varphi }_0^2}\, ,
    }
    \eq{
        c_{\psi} \equiv \frac{4 (-1+3 \lambda ) \dot{\varphi }_0^2}{8 H^2 (-1+3 \lambda )+\kappa ^2 (-1+\lambda ) \dot{\varphi }_0^2}\, ,
    }
    \eq{
        c_{\psi\varphi} \equiv \frac{8 H (-1+3 \lambda ) \dot{\varphi }_0}{8 H^2 (-1+3 \lambda )+\kappa ^2 (-1+\lambda ) \dot{\varphi }_0^2}\,
,
    }

    \eq{
        f_{\varphi} \equiv \frac{\kappa ^2 \dot{\varphi }_0 \left((H-3 H \lambda ) \dot{\varphi }_0
-(-1+\lambda ) V'\right)}{8 H^2 (-1+3 \lambda )+\kappa ^2
(-1+\lambda ) \dot{\varphi }_0^2}\, ,
    }
    \eq{
        f_{\psi} \equiv \frac{12 H (1-3 \lambda )}{\kappa ^2}\, ,
    }
    \eq{
        f_{\psi\varphi} \equiv -3 \dot{\varphi }_0^2\, ,
    }
    \eq{
        \tilde{f}_{\psi\varphi} \equiv \frac{(-1+3 \lambda ) \left(-\kappa ^2 \dot{\varphi }_0^3+8 H V'\right)}{8 H^2 (-1+3 \lambda )
+\kappa ^2 (-1+\lambda ) \dot{\varphi }_0^2}\, ,
    }

    \eq{
        h_{\psi} \equiv \frac{4 H \kappa ^2 \Lambda  \mu ^2}{8 H^2 (-1+3 \lambda )+\kappa ^2 (-1+\lambda ) \dot{\varphi }_0^2}\, ,
    }
    \eq{
        h_{\psi\varphi} = -\frac{\kappa ^4 (-1+\lambda ) \Lambda  \mu ^2 \dot{\varphi }_0 }{16 H^2 (1-3 \lambda )^2
+2 \kappa ^2 (-1+\lambda ) (-1+3 \lambda ) \dot{\varphi }_0^2}\, ,
    }

    \eq{
        \omega_{\psi} \equiv \frac{\kappa ^2 \Lambda  \mu ^2}{-4+12 \lambda }\, ,
    }
    \eq{
        \omega_{\varphi\psi} \equiv -\frac{\kappa ^4 \Lambda  \mu ^2 \left(H (-1+3 \lambda ) \dot{\varphi }_0+(-1+\lambda ) V'\right)}{2 (-1+3 \lambda ) \left(8 H^2 (-1+3 \lambda )+
\kappa ^2 (-1+\lambda ) \dot{\varphi }_0^2\right)}\, ,
    }

    \eq{
        d_{\psi} \equiv -\frac{\kappa ^2 (-1+\lambda ) \mu ^2 \left(16 H^2 (1-3 \lambda )^2+\kappa ^4 \Lambda ^2 \mu ^2+2 \kappa ^2 \left(1-4 \lambda +3 \lambda ^2\right) \dot{\varphi }_0^2\right)}{8 (1-3 \lambda )^2 \left(8 H^2 (-1+3 \lambda )+\kappa ^2 (-1+\lambda ) \dot{\varphi
        }_0^2\right)} \,,
    }

    \eq{
        m_{\varphi} \equiv \frac{\kappa ^4 \dot{\varphi }_0^4+8 H \kappa ^2 (1-3 \lambda ) \dot{\varphi }_0 V'-4 \kappa ^2 (-1+\lambda ) \left(V'\right)^2+32 H^2 (1-3 \lambda ) V''-4 \kappa ^2 \dot{\varphi }_0^2 \left(3 H^2 (-1+3 \lambda )+(-1+\lambda ) V''\right)}{8 \left(8 H^2 (-1+3 \lambda )+\kappa ^2 (-1+\lambda ) \dot{\varphi
        }_0^2\right)} \,,
    }
    \eq{
        m_{\psi} = \frac{3 \left(12 H^2 (1-3 \lambda )+\kappa ^2 \dot{\varphi }_0^2\right)}{2 \kappa
        ^2} \,,
    }
    \eq{
        m_{\psi\varphi} = 3V' \,.
    }

\subsection{Coefficients in (\ref{2nd_action_zeta_psi})}\label{app_coeff_2}

    \eq{
        c_{\zeta} \equiv \frac{4 (-1+3 \lambda )  \dot{\varphi }_0^2}{8 H^2 (-1+3 \lambda )+\kappa ^2 (-1+\lambda ) \dot{\varphi }_0^2}\, ,
    }

    \eq{
        f_{\zeta} \equiv - \frac{ \dot{\varphi }_0 \left(3 H \dot{\varphi }_0+V'\right)}{H^2}\, ,
    }
    \eq{
        f_{\psi} \equiv \frac{12 H (1-3 \lambda )}{\kappa ^2}+\frac{3 \dot{\varphi }_0^2}{H}-\frac{\dot{\varphi }_0 V'}{H^2}\, ,
    }
    \eq{
        f_{\zeta\psi} \equiv - \frac{ \dot{\varphi }_0 V'}{H^2}\, ,
    }
    \eq{
        \tilde{f}_{\zeta\psi} \equiv - \frac{ \dot{\varphi }_0 V'}{H^2}\, ,
    }

    \eq{
        h_{\psi} \equiv -\frac{ \kappa ^2 \Lambda  \mu ^2 }{2 H-6 H \lambda } \,,
    }
    \eq{
        h_{\zeta\psi} \equiv \frac{  \kappa ^4 (-1+\lambda ) \Lambda  \mu ^2 \dot{\varphi }_0^2}{2 H (-1+3 \lambda )
\left(8 H^2 (-1+3 \lambda )+\kappa ^2 (-1+\lambda ) \dot{\varphi
}_0^2\right)}\, ,
    }

    \eq{
        d_{\zeta} \equiv \frac{ g_2 \dot{\varphi }_0^2}{H^2} \,,
    }
    \eq{
        \tilde{d}_{\zeta} \equiv \frac{ g_3 \dot{\varphi }_0^2}{H^2}\, ,
    }
    \eq{
        \tilde{d}_{\psi} \equiv \frac{ g_3 \dot{\varphi }_0^2}{H^2}\, ,
    }
    \eq{
        d_{\psi} \equiv \frac{g_2 \dot{\varphi }_0^2}{H^2}-\frac{\kappa ^2 (-1+\lambda )
\mu ^2 \left(16 H^2 (1-3 \lambda )^2+\kappa ^4 \Lambda ^2 \mu ^2+2
\kappa ^2 \left(1-4 \lambda +3 \lambda ^2\right) \dot{\varphi
}_0^2\right)}{8 (1-3 \lambda )^2 \left(8 H^2 (-1+3 \lambda )+\kappa
^2 (-1+\lambda ) \dot{\varphi }_0^2\right)}\, ,
    }
    \eq{
        d_{\zeta\psi} \equiv \frac{2  g_2 \dot{\varphi }_0^2}{H^2}\, ,
    }
    \eq{
        \tilde{d}_{\zeta\psi} \equiv \frac{2 g_3 \dot{\varphi }_0^2}{H^2}\, ,
    }

    \ea{
        \omega_{\psi} &\equiv \frac{  2 H^2 \kappa ^2 (-1+3 \lambda ) \Lambda
\mu ^2+\left(\kappa ^4 \Lambda  \mu ^2-8 (1-3 \lambda )^2 g_1\right)
\dot{\varphi }_0^2 }{8 (H-3 H \lambda )^2}\, ,
    }
    \eq{
        \omega_{\zeta} \equiv -\frac{ g_1 \dot{\varphi }_0^2}{H^2}\, ,
    }
    \eq{
        \omega_{\psi\zeta} \equiv -\frac{  g_1 \dot{\varphi }_0^2}{H^2}\, ,
    }
    \eq{
        \tilde{\omega}_{\zeta\psi} \equiv \frac{  \left(\kappa ^4 \Lambda  \mu ^2-8 (1-3 \lambda )^2 g_1\right) \dot{\varphi }_0^2}{8
(H-3 H \lambda )^2}\, ,
    }

    \ea{
        m_{\zeta} &\equiv \frac{ \left(-3 H \kappa ^2 \dot{\varphi }_0^4+24 H^2 (-1+3 \lambda )
\dot{\varphi }_0 V'-2 \kappa ^2 \dot{\varphi }_0^3 V'+4 H (-1+3
\lambda ) \left(V'\right)^2+4 H (-1+3 \lambda ) \dot{\varphi }_0^2
\left(9 H^2-V''\right)\right)}{8 H^3 (-1+3 \lambda )}\, ,
    }
    \ea{
        m_{\psi} &\equiv  \frac{1}{8 H^3 \kappa ^2 (-1+3 \lambda )} \left\{   3 H \kappa ^4 \dot{\varphi }_0^4+24 H^2 \kappa ^2 (1-3 \lambda ) \dot{\varphi }_0 V'-2 \kappa ^4 \dot{\varphi }_0^3 V'  \right. \\
        &\qquad\qquad \left. -4 H (1-3 \lambda ) \left(36 H^4 (1-3 \lambda )+\kappa ^2 \left(V'\right)^2\right)-
4 H \kappa ^2 (-1+3 \lambda ) \dot{\varphi }_0^2 \left(6
H^2+V''\right)   \right\}   \,,
    }
    \ea{
        m_{\zeta\psi} &\equiv \frac{  \left(-\kappa ^2 \dot{\varphi }_0^3 V'+2
H (-1+3 \lambda ) \left(V'\right)^2+2 H (1-3 \lambda ) \dot{\varphi
}_0^2 V''\right)}{2 H^3 (-1+3 \lambda )}\, .
    }

\subsection{$c_s^2$ and $M^2$ in (\ref{eom_UV})}{\label{app_mass}}

The exact form for the ``effective speed of sound" $c_s^2$ is given by
    \ea{
        c_s^2 &\equiv \frac{1}{2048 H^2 (1-3 \lambda )^6 g_3 \dot{\varphi }_0^4} \\
        &\qquad\times \left\{ H^2 \kappa ^4 (-1+\lambda )^2 \mu ^4 \left(16 H^2 (1-3 \lambda )^2+\kappa ^4 \Lambda ^2 \mu ^2\right)^2 \right. \\
        &\qquad\qquad  +2 \dot{\varphi }_0^2 \left[ 16 H^4 \kappa ^6 \left(1-4 \lambda +3 \lambda ^2\right)^3 \mu ^4+H^2 \kappa ^{10} (-1+\lambda )^3 (-1+3 \lambda ) \Lambda ^2 \mu ^6 \right.\\
        &\qquad \left.\left. +32 (-1+3 \lambda )^3 g_3 \left(16 H^4 \kappa ^2 (1-3 \lambda )^2 \Lambda  \mu ^2+\left(\kappa ^4 \Lambda  \mu ^2-8 (1-3 \lambda )^2 g_1\right) \dot{\varphi }_0^2 \left(8 H^2 (-1+3 \lambda )+\kappa ^2 (-1+\lambda ) \dot{\varphi
        }_0^2\right)\right)\right]\right\} \,.
    }
In the case $\lambda=1$,
    \eq{
        c_s^2 = \frac{64 H^4 \kappa ^2 \Lambda  \mu ^2+16 H^2 \left(\kappa ^4 \Lambda  \mu ^2-32 g_1\right) \dot{\varphi }_0^2}{256 H^2 \dot{\varphi
        }_0^2} \,,
    }
which reduces further to $c_s^2 = -2g_1$ if we set $\Lambda=0$, as
expected.

 The effective mass-square term $M^2$ is given by
\ea{
    M^2 &= \frac{1}{256 c_{\zeta }^4 \tilde{d}_{\psi }^4}\left\{ h_{\zeta \psi }^6 \left(\tilde{d}_{\zeta \psi }^2-4 \tilde{d}_{\zeta } \tilde{d}_{\psi }\right)-4 c_{\zeta } h_{\zeta \psi }^4 \left(2 \tilde{d}_{\psi } \left(-d_{\zeta \psi } \tilde{d}_{\zeta \psi }+2 d_{\zeta } \tilde{d}_{\psi }\right)+d_{\psi } \left(3 \tilde{d}_{\zeta \psi }^2-8 \tilde{d}_{\zeta } \tilde{d}_{\psi }\right)\right) \right. \\
    &\qquad + 16 c_{\zeta }^2 h_{\zeta \psi }^2 \left[ 4 d_{\psi } \tilde{d}_{\psi } \left(-d_{\zeta \psi } \tilde{d}_{\zeta \psi }+d_{\zeta } \tilde{d}_{\psi }\right)+ d_{\psi }^2 \left(3 \tilde{d}_{\zeta \psi }^2-4 \tilde{d}_{\zeta } \tilde{d}_{\psi }\right) \right. \\
    &\qquad\qquad\qquad\qquad \left. + \tilde{d}_{\psi } \left(H h_{\psi } \left(\tilde{d}_{\zeta \psi }^2-2 \tilde{d}_{\zeta } \tilde{d}_{\psi }\right)  +\tilde{d}_{\psi } \left(d_{\zeta \psi }^2-H h_{\zeta \psi } \tilde{d}_{\zeta \psi }+2 \omega _{\psi \zeta } \tilde{d}_{\zeta \psi }-4 \omega _{\zeta } \tilde{d}_{\psi }+2 \tilde{d}_{\zeta \psi } \tilde{\omega }_{\zeta \psi }\right)\right)\right] \\
    &\qquad  -32 c_{\zeta }^3 \left[2 d_{\psi }^3 \tilde{d}_{\zeta \psi }^2-4 d_{\zeta \psi } d_{\psi }^2 \tilde{d}_{\zeta \psi } \tilde{d}_{\psi }+2 d_{\psi } \tilde{d}_{\psi } \left(H h_{\psi } \tilde{d}_{\zeta \psi }^2+\tilde{d}_{\psi } \left(d_{\zeta \psi }^2+\tilde{d}_{\zeta \psi } \left(-3 H h_{\zeta \psi }+2 \left(\omega _{\psi \zeta }+\tilde{\omega }_{\zeta \psi }\right)\right)\right)\right) \right. \\
    &\qquad\qquad  -\tilde{d}_{\psi }^2 \left(3 H f_{\psi } \tilde{d}_{\zeta \psi }^2-2
\left(m_{\psi } \tilde{d}_{\zeta \psi }^2-2 \tilde{d}_{\psi }
(m_{\zeta \psi } \tilde{d}_{\zeta \psi }+3 H f_{\zeta }
\tilde{d}_{\psi }-2 m_{\zeta } \tilde{d}_{\psi }-3 H
\tilde{d}_{\zeta \psi } \tilde{f}_{\zeta \psi } ) \right) \right. \\
&\qquad\qquad\qquad\qquad \left.\left.\left. +2 d_{\zeta \psi }
\left(H h_{\psi } \tilde{d}_{\zeta \psi }+\tilde{d}_{\psi } \left(-3
H h_{\zeta \psi }+2 \left(\omega _{\psi \zeta }+\tilde{\omega
}_{\zeta \psi }\right)\right)\right)\right)\right]\right\} \,.
    }

\subsection{$\tilde{c}_s^2$ and $\tilde{M}^2$ in (\ref{eom_IR})
}\label{appsec_cs_mass_IR}

\ea{{\label{app_tilde_cs}}
    \tilde{c}_s^2 &\equiv -\frac{1}{2 c_{\zeta } \left(3 H f_{\psi }-2 m_{\psi }\right){}^2 } \times \left\{ H h_{\psi } \left(m_{\zeta \psi }-3 H
\tilde{f}_{\zeta \psi }\right)^2 \right. \\
&\qquad\qquad \left. +\left(3 H f_{\psi }-2 m_{\psi }\right)
\left[-6 H f_{\psi } \omega _{\zeta }+4 m_{\psi } \omega _{\zeta }+H
h_{\zeta \psi } \left(m_{\zeta \psi }-3 H \tilde{f}_{\zeta \psi
}\right)-2 \left(m_{\zeta \psi }-3 H \tilde{f}_{\zeta \psi }\right)
\left(\omega _{\psi \zeta }+\tilde{\omega }_{\zeta \psi
    }\right)\right] \right\} \,,
} and \eq{
    \tilde{M}^2 \equiv -\frac{-9 H^2 f_{\zeta } f_{\psi }+6 H f_{\psi } m_{\zeta }+m_{\zeta \psi }^2+6 H f_{\zeta } m_{\psi }-4 m_{\zeta } m_{\psi }-6 H m_{\zeta \psi } \tilde{f}_{\zeta \psi }+9 H^2 \tilde{f}_{\zeta \psi }^2}{6 H c_{\zeta } f_{\psi }-4 c_{\zeta } m_{\psi
    }} \,,
} with parameters are given in Appendix \ref{app_coeff_2}.

\end{document}